# Investigation of DNA denaturation in Peyrard-Bishop-Dauxois model by molecular dynamics method

Likhachev I.V., Lakhno V.D.
*The Institute of Mathematical Problems of Biology, Russian Academy of Sciences - a branch of the KIAM RAS.*
*IMPB RAS, 1, Professor Vitkevich St., 142290, Pushchino, Moscow Region, Russia*
*e-mail address: ilya_lihachev@mail.ru*

The phase transition of $(PolyA/PolyT)_{100}$ duplex into the denaturated state is studied in the Peyrard-Bishop-Dauxois model by the method of direct molecular-dynamical modeling. The temperature dependencies of the total energy and heat capacity of the duplex are calculated. The approach applied can be used to calculate the statistical properties of the duplexes of any length and nucleotide composition.

In modeling molecular dynamics of biomacromolecules use is made of interatomic potentials which are always limited in value in the case of large distances between the atoms (Lennard-Jones, Coulomb, Morse, etc. [1]). This involves some problems when calculating the thermodynamic properties of biomacromolecules. Indeed, the statistical sum always diverges for bounded potentials. This means that macromolecules should always be in denaturated state at arbitrarily low temperature. However actually this is not the case since the concentration of molecules in a solution is finite which determines a characteristic cutoff scale in the calculations of the statistical sum.

If we formally deal not with an ensemble, but with a single molecule in an infinite solvent and the temperature is much lower than the denaturation temperature then potential energy always reaches a stationary quasi-equilibrium state which exists far longer than any experiment concerned. This equilibrium state will naturally coincide with the state which we would obtain dealing with an ensemble of molecules.

Hence the problem of modeling is not a formal calculation of the statistical sum over infinite time corresponding to the decomposed system, but the calculation leading to just the quasi-equilibrium state and finding of its thermodynamic characteristics.

A problem, however arises when the temperature appears to be close to the value for which two characteristic times become equal (i.e. the time during which potential energy reaches the stationary state and the time of the system's decomposition).

This situation arises, for example, in calculations of the thermodynamic characteristics of DNA whose decomposition temperature not far exceeds the room one. In this case a direct calculation of the statistical sum is impossible, since it leads to a decomposition of the molecule during a finite time.

This paper just deals with consideration of this problem. We describe a method which can be effectively used to calculate the thermodynamic characteristics of the molecule in the vicinity of the transition and the transition *per se*.

Melting (denaturation) of DNA [1] occurs as a result of amplification of temperature fluctuations. In this case both the chains move away from each other breaking the hydrogen bonds. The chains *per se* remain nonseparable.

A popular dynamical model of DNA describing such behavior of the duplex – the Peyrard-Bishop-Dauxois (PBD) model – simulates both the behavior of breaking hydrogen bonds (Peyrard-Bishop model) with the use of the Morse potential and the inseparability of each chain due to Peyrard-Dauxois addition [2].

In this model a chain of DNA nucleotides is presented as a system of material points in a one-dimensional space whose motion is described by classical motion equations:

$$m_i \frac{d^2 x_i}{dt^2} = -\frac{dU}{dx_i}$$

where i=1..N; N is the number of particles; m is the reduced mass of a particle; x is the coordinate of deviation from the equilibrium position between the nucleotides; U is the potential; t is the time.

According to the PBD model, each particle occurs in a potential field:

$$U = U_{Morze} + W$$

In the PBD model the interaction of the nucleotides in a pair is described by the Morse potential:

$$U_{Morze}(x_i) = D(1 - e^{-\alpha x_i})^2$$

The interaction of neighboring pairs has the form:

$$W(x_i, x_{i-1}) = \frac{k}{2}\left(1 + \rho e^{-a(x_i + x_{i-1})}\right)(x_i - x_{i-1})^2$$

Consideration is given to a chain of 100 AT base pairs for the parameter values: D=k= 0.92 kcal/mole, $\alpha$=4.45 $Å^{-1}$, a=0.35 $Å^{-1}$, $\rho$=0.5. m=300 a.m.u, used in [2].

In order to integrate the motion equations in the PBD model we will use the velocity Verlet algorithm [3].

The temperature is taken into account in the model in the same manner as in the method of all-atom molecular dynamics [4], with the use of a collisional thermostat, where the medium in which the modeled system occurs is simulated by point particles demonstrating the Maxwell velocity distribution [5, 6, 7]. The velocity distribution corresponds to a certain temperature $T_{ref}$. At random times virtual particles of the medium elastically collide with the particles of the system. The motion equations have the form:

$$m_i \frac{dv_i}{dt} = F_i + \sum_k f_{ik} \cdot \delta(t - t_{ik})$$

$$F_i = -\frac{dU}{dx_i}$$

Here $\delta(t)$ is the Dirac delta function, $f_{ik}$ is a stochastic source of force leading to jumps of the i-th atom velocity at random times $t_{ik}$. The value of the velocity jump is calculated as a result of collision of two point particles which had the velocities v (for a modeled particle) and $v_0$ (for a virtual particle) before the collision:

$$\Delta v(t) = \frac{2m_0}{m_0 - m}\left(v_0(t) - v(t)\right)$$

Here m is the mass of a modeled particle, $m_0$ is the mass of a virtual particle. The velocities $v_0$ are chosen from the Maxwell distribution.

Collisions take place in accordance with the Poisson process which is determined by the only parameter $\lambda_0$ – the mean value of the collisions of an atom with medium particles per a unit of time, i.e. the frequency of collisions.

In the cource of modeling the temperature of the collisional thermostat $T_{current}$ changed linearly at a constant velocity $V_t$=0,1 K/ns, $T_{current}=T_{initial}+V_t*t$; which yields the graph of the dependence of the specific potential, kinetic and full energy on the time (see Fig. 1).

The initial temperature is chosen to be demonstrably less than the lowest point of DNA denaturation. In our example $T_{initial}$=250 K.

The velocity of heating per a unit time is rather small so that we can believe that on any rather small time interval the system occurs in equilibrium. According to numerical experiments, the system's relaxation – leveling off of the potential energy from temperature 0 K to 250 K – occurs for less than 1 ns.

Fig. 1 illustrates the temperature dependence of the energy of a polynucleotide chain consisting of 100 base pairs in a single computational experiment (single realization). The kinetic energy changes linearly in the range from 300 K to 400 K. The potential energy and, as a consequence, the full energy have two near-linear regions and a transition region where the energy changes sharply. The time-averaged kinetic energy depends linearly on time due to the collisional thermostat.

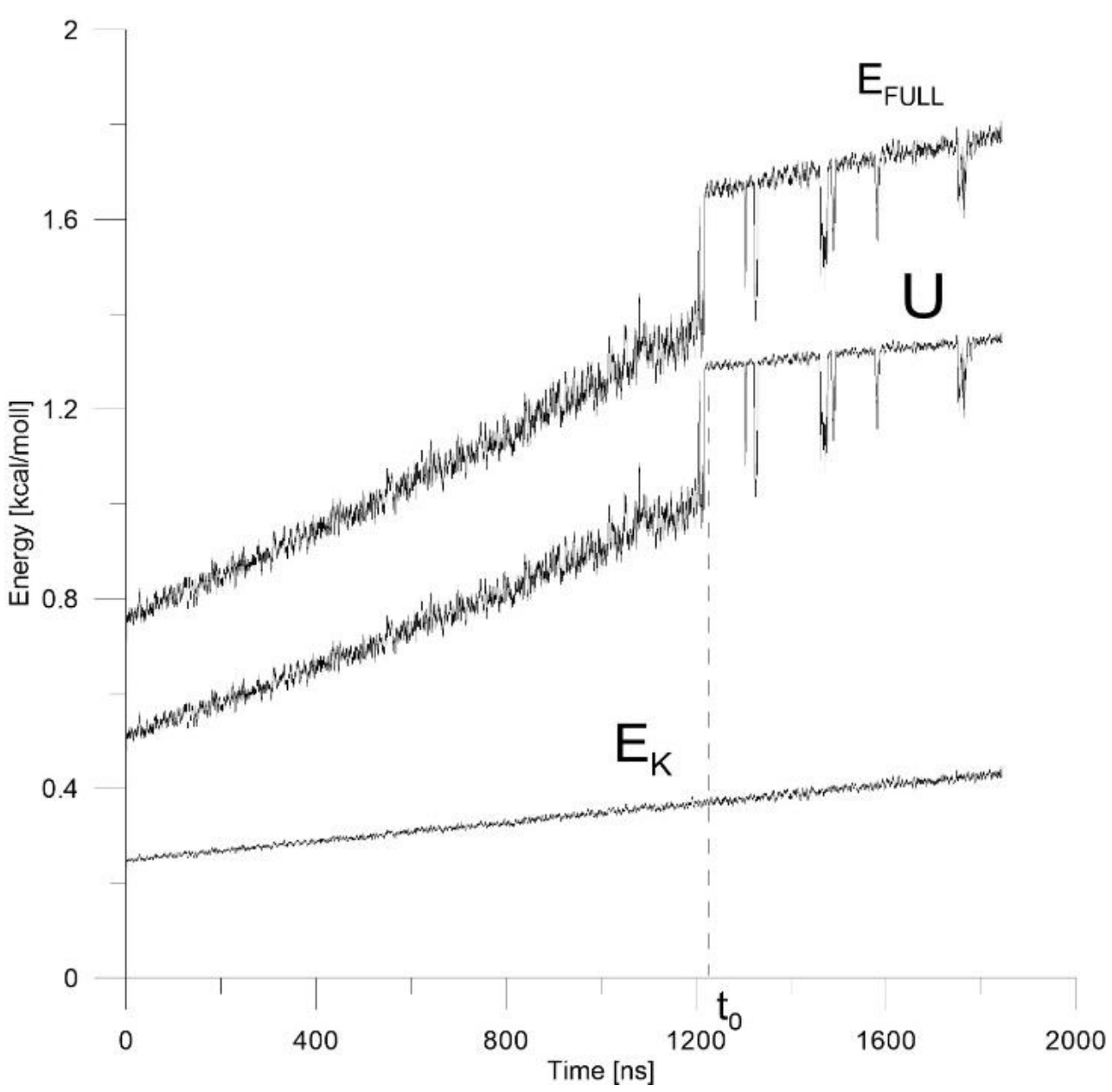

*FIG. 1. Time dependence of the specific full ($E_{FULL}$) potential (U) and kinetic ($E_K$) energies for a single realization in the numerical experiment as the thermostat temperature linearly increases $T_{REF}$=250 К+0.1 К/ns*t.*

The time dependence of the full and potential energy shown in Fig. 1 describes the transition from the quasi-harmonic regime to the denaturated one. The time $t_0$ corresponds to the thermostat temperature $T_{REF}$=250 К+0.1 К/нс*$t_0$. Avaraging of these times over realizations yields the transition temperature (see Fig. 2).

Actually Fig. 1 describes the energy gap whose order of magnitude is $k_BT$, where T is the temperature corresponding to the time of the jump onset $t_0$. It separates the denaturated state (to the right of the jump) from the non-denaturated one (to the left of the jump). In different realizations the jump occurs at different times since the external force (thermostat) acts randomly. The early transition illustrates a larger energy gap, the late transition – a smaller one. The availability of such a gap, as was mentioned above, is caused by the divergence of the statistical sum for bounded potentials which presents computational difficulties for the description of the melting process.

To analyze the macroscopic properties such as the heat capacity, averaging over a large number of the systems (realizations) is required.

Fig. 2 illustrates the temperature dependence of the specific full energy E(T) averaged over N=512 realizations which accounts for one site of a chain of 100 nucleotide pairs.

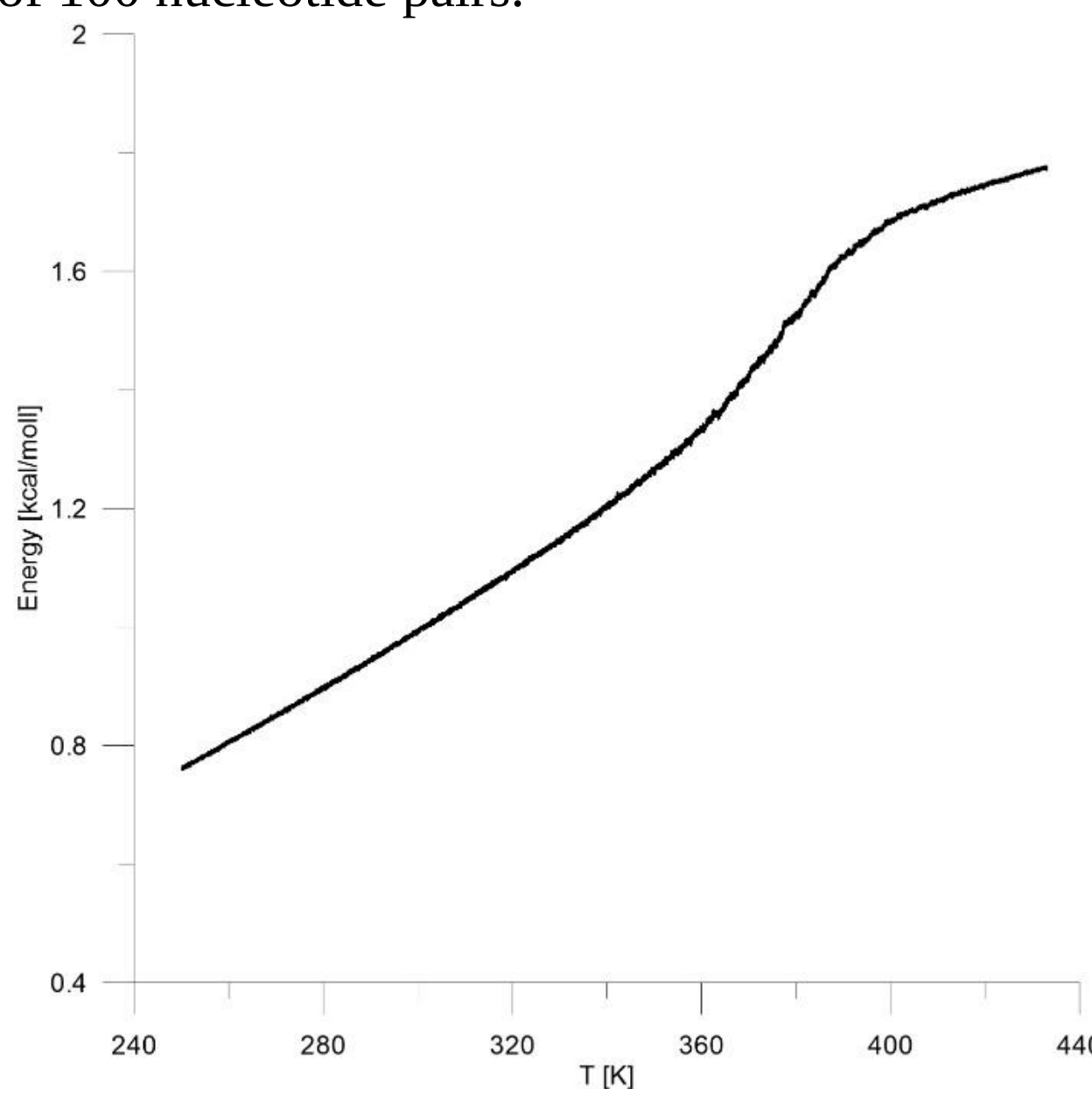

*FIG. 2. Temperature dependence of the specific full energy averaged over 512 realizations*

Fig. 3 illustrates the temperature dependence of a specific heat capacity $C_V = \frac{dE}{dT}$ for four different systems averaged over 512 realizations: a system of 25 particles (ρ=0.5), a system of 50 particles (ρ=0.5), a system of 100 particles (ρ=0.5) and a system of 100 particles (ρ=2). In order to find a derivative, the full energy function was approximated by two straight lines (before and after the transition) and by the polynomial of degree 5 in the vicinity of the transition. After approximation the derivative was found analytically. The boundaries of the approximation region were found by the minimization method so that the piecewise analytical function would have the smallest squared deviation from the data of modeling.

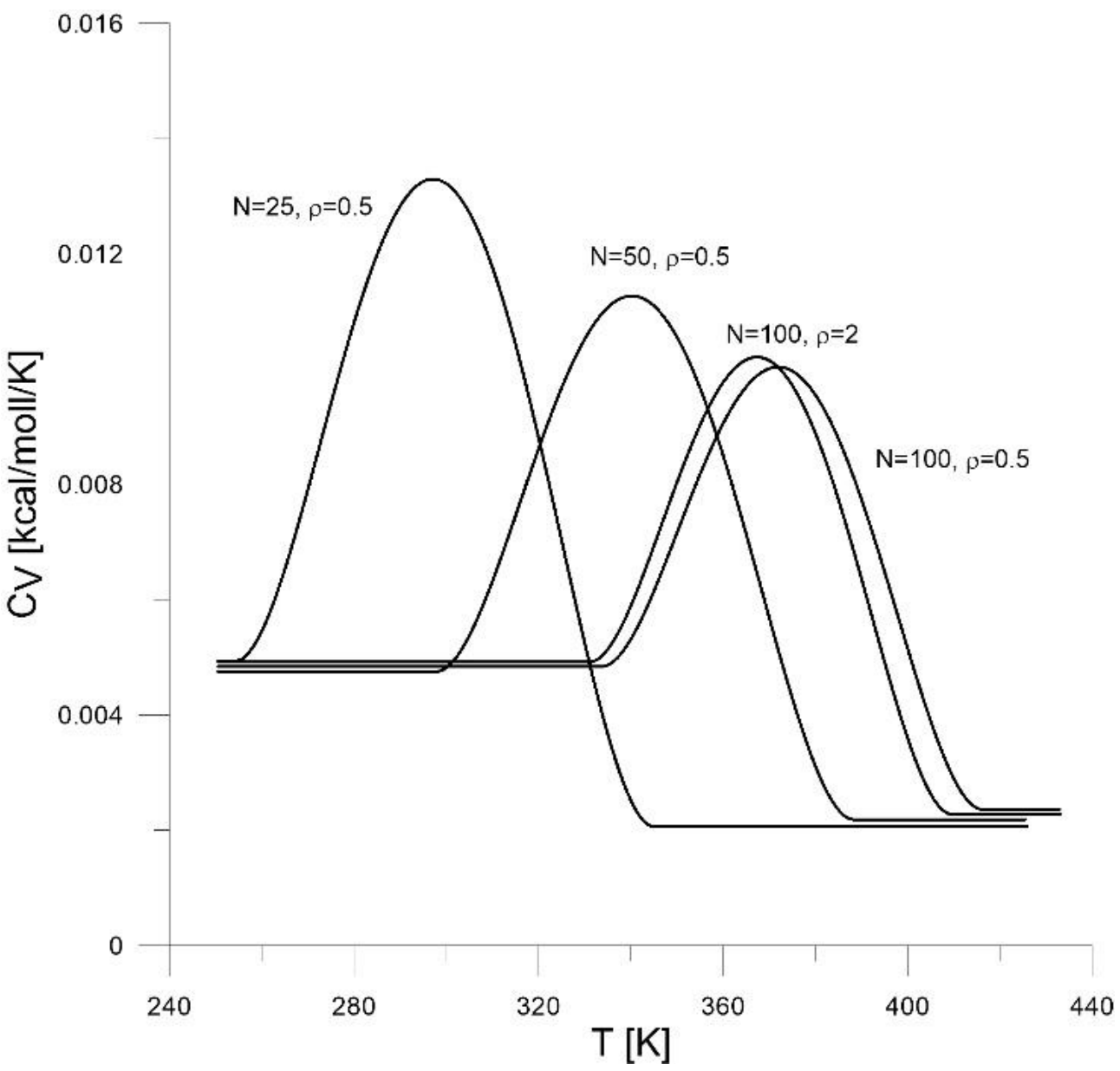

*FIG. 3. Specific heat capacity as a derivative of the full energy for four systems.*

The curves in Fig. 3 describe the process of DNA melting which corresponds to a wide peak of the heat capacity. The value of the heat capacity to the left of the peak is larger than that to the right of the peak since on the left, the heat supplied to the chain is spent on the enhancement of the potential energy between neighboring Watson-Crick pairs and on the enhancement of the potential energy of the chain plus the kinetic energy. To the right of the peak when the chain is completely denaturated its further heating is not spent on the potential energy between the nucleotides in pair. Numerically, the value to the left of the peak of the heat capacity exceeds two universal gas constants (2R), and to the right of the peak tends to R. Prior to the peak, we have two potentials close to parabolic: the Morse potential at low temperatures is close to parabolic, and the stacking interaction potential is close to parabolic due to the tendency of the exponent to zero.

The temperature at which the peak is maximum corresponds to the melting point of DNA. At this point the full energy of a homogeneous duplex is equal to ND, where D is a constant of the Morse potential.

In papers [2, 8, 9], as distinct from Fig. 3 and paper [10], stepwise curves were obtained for the heat capacity. In the region of near-critical temperatures in the PBD model the kinetic energy obtained from the thermostat is spent on the enchancement of the potential energy of internucleotide interaction. This explains the smooth decrease of the heat capacity after the peak. The transfer integral technique used in [2] to calculate $C_V(T)$ yields a δ-shaped peak which corresponds to the results of modeling of very long chains. As distinct from the transfer integral technique, our approach can be used for the chains of any length and any nucleotide composition. Besides the transfer integral technique is limited by the Peyrard-Bishop model. Even simplest sophistication of the model, such as consideration of the PBD model, does not allow any analytical solutions of relevant integral equations which renders calculation of actual systems impossible.

In paper [11] the thermodynamical functions were calculated by the direct method for calculation of the statistical sum as a multidimensional integral. As a rule, this method is essentially limited by the chain length. The most serious limitation of the method, however, lies in the fact that it enables investigations of only statistical thermodynamic states, while our method makes possible investigations of dynamical and statistical properties of the system.

The peak of the heat capacity received by us depends on the type of a nucleotide sequence. This conclusion is consistent with paper [12]. When the number of physical particles is smaller the phase transition occurs at lower temperature. This conclusion is consistent with paper [13] which deals with a slightly modified model which is similar to our model in the region of small oscillations.

Comparison of the results obtained by our method and those obtained by TI-method from paper [2] is given in Fig. 4 which illustrates the temperature dependence of the mean coordinate value of a nucleotide pair displacement from the equilibrium position over the realizations. The extent of coincidence is rather high which suggests the adequacy of the calculations performed. Some distinctions can be explained by the fact that in [2] use was made of the Nose-Hoover thermostat while we used a collisional thermostat.

The thermostats by Berendsen [14], Nose and Nose-Hoover [15] can introduce some artifacts in the investigations. All these thermostats set the mean temperature well. However as one very hot (or very cold) particle emerges these thermostats in an effort to maintain the mean temperature cool off (or heat) the system which occurs at a temperature close to a certan predetermined value. After some time the temperatures of all the particles of the system become balanced due to the interparticle interaction but not due to the thermostat. As regards the Morse potential, cold particles can emerge while attempting to get out of the potential well. If a particle does not fall back to the well its velocity is lower than that in a quasi-harmonic regime.

As distinct from the Berendsen and Nose-Hoover thermostats, the collisional thermostat acts on each particle individually irrespective of the temperature of the rest of the particles in the system which is more consistent with reality.

Refering to paper [2] we notice that our mean coordinate values coincide with their results, while the heat capacity does not. The reason is that in paper [2] the coordinates are calculated by the molecular dynamics method until the moment of denaturation and the heat capacity is given only by the transfer integral technique. We use the molecular dynamics method to calculate the coordinates both before the denaturation and after it.

The results obtained seem to be rather general and can be generalized to other systems. In particular, a wide peak of the heat capacity was observed experimentally in proteins [16].

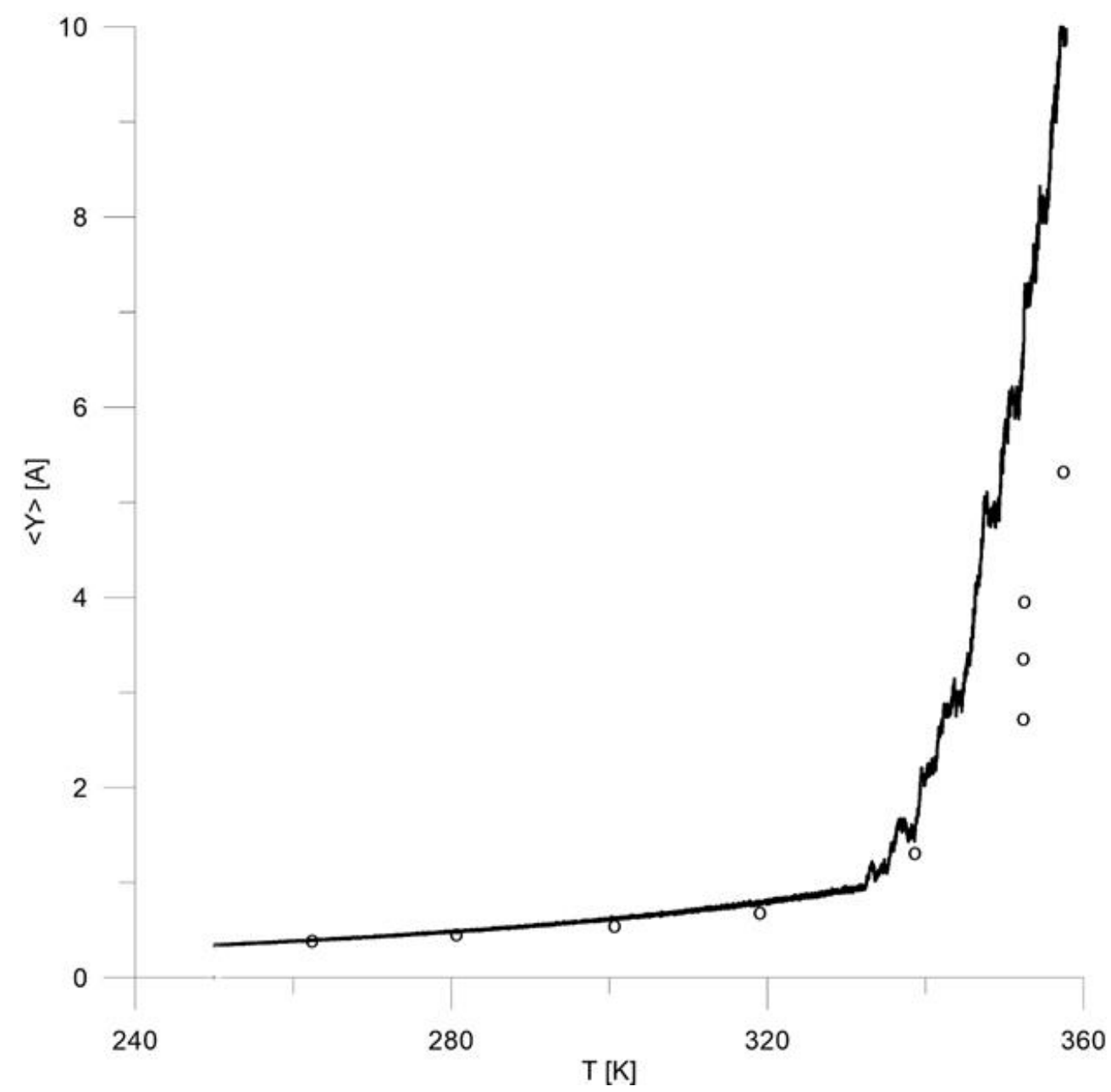

*FIG. 4. Temperature dependence of the mean coordinate value (black curve). The length of the chain is 100 nucleotides, the number of realizations is 100. The circles represent the data of molecular modeling with the use of the Hose-Hoover thermostat from paper [2].*

## ACKNOWLEDGEMENTS

The authors express their gratitude to N.K. Balabaev for his proposals on the technique of carrying out the experiments and to N.S Fialko for her critical remarks.

The research was carried out using the equipment of the shared research facilities of HPC computing resources at Lomonosov Moscow State University [17] and supercomputers k100 and k60 of the KIAM RAS.